\documentstyle[12pt]{article}

\begin{document}
\begin{titlepage}
\begin{center}
\today     \hfill    ACT-16/94 \\
          \hfill    hep-ph/9412262 \\

\vskip .5in

{\large \bf
Supersymmetry Breaking Scheme and  The Derivation of $M_{GUT}=10^{16}GeV$
  from A String Model} \\
{\bf Rulin Xiu} \\

{\em Lyman Laboratory of Physics \\
    Harvard University\\
      Cambridge, MA 02138}

\end{center}

\vskip .5in

\begin{abstract}
We propose that a  modified version of string-inspired supersymmetry
breaking schemes in \cite{wit2} and \cite{ross} may make affine level one
SU(5) and SO(10) string models with intermediate gauge symmetry breaking scale
$M_{GUT}=10^{16}GeV$ possible; doublet-triplet problem can also be solved.
In particular, we propose that some charged gauge background  VEVs
might be turned on when gaugino condensation happens in the hidden sector
and would break  gauge symmetry. One can dynamically determine
gaugino condensation scale, local supersymmetry breaking scale and
intermediate gauge symmetry breaking scale  $M_{GUT}$
 in terms of dilaton VEV. With dilaton VEV, $\langle S
 \rangle \sim 2$, and $E_6$ hidden sector gaugino condensation,
we can obtain $M_{GUT}=10^{16}GeV$. We also discuss how to generate the
mass hierarchy within the framework of this supersymmetry breaking scheme.
We show that
the large mass hierarchy restricts string models to those that have a
no-scale structure and in which the gauge coupling function does not receive
string threshold corrections.

\end{abstract}
\end{titlepage}

The precision weak scale measurement indicates that the minimal supersymmetric
grand unification model leads to a good agreement with a single unification
scale of $M_{GUT}=10^{16\pm0.3}GeV$\cite{gut1} and a best fit for
supersymmetry breaking scale $M_{SUSY}$ around 1 TeV \cite{gut1,gut2,gut3}.
It is amazing that the
supersymmetry grand unification idea works perfectly in this respect and
predicts correctly the present experimental value for $sin^2\theta_W(m_Z)$,
once one sets $M_{SUSY}$ around the TeV scale and use the actual values
for $\alpha_s(M_Z)$ and $\alpha_{em}(M_Z)$. In addition to this success,
the bottom to tau quark mass ratio is also predicted correctly; the proton
lifetime is predicted to be above the present experimental limits.

String theory is so far the leading candidate for a consistent framework
for unifying the standard model and gravity \cite{string1,string2}.
In the framework of
string theory, different mass scales are ultimately determined by the vacuum
expectation values of dilaton and moduli fields, which in turn should
be determined by the string dynamics. One would naturally ask whether
it is possible to derive  $M_{GUT}=10^{16}GeV$ and $M_{SUSY}=1 TeV$
from string phenomenology. To answer this question, we need first
have  a brief review of gauge coupling constant unification in the string
phenomenology.

Whereas the unification of the running gauge coupling constants is a
coincidence and the gauge unification scale is arbitrary in the standard model,
a relation among gravitational, gauge coupling constants, and the ``gauge
unification scale''  is required in string theory. More specifically,
this relation has been derived \cite{kap,tom} to be:
$g^2_a(\frac{1}{2}g^2M_{pl})
=g^2/k_a-\epsilon_a/4\pi$, where $M_{Pl}=1/\sqrt{8\pi G_N}$ is the reduced
 Planck mass
($G_N$ is the Newton's constant),  $\epsilon_a$ is the string threshold
corrections, and $k_a$ is the affine level. In the simple case,
$k_a=1$ and $\epsilon_a=0$, one obtains the gauge coupling constant unification
but at the scale around $10^{17}\sim 10^{18} GeV$. To confront string theory
with the weak scale gauge coupling constant measurements, extensive discussion
has been made on  Standard-Model-like-string-models
\cite{extra,thre1,thre2,high} with extra heavy
fermions, string threshold corrections, or of higher affine levels, but none
 of which  lead to gauge coupling constant unification at
$M_{GUT}=10^{16}GeV$.

In this letter, we propose that a modified version of the supersymmetry
breaking scheme proposed in \cite{wit2,ross} may yield $M_{GUT}=10^{16}GeV$ in
string models. In particular, we propose that some charged background field
VEVs in  the observable $E_8$ sector
might be turned on when gaugino condensation happens in the hidden sector
and would break  gauge symmetry. We point out that this supersymmetry
breaking scheme makes the affine level one SU(5) or SO(10) string models
with intermediate gauge symmetry breaking to the standard model possible.
We also discuss
about generating the large mass hierarchy between  $M_{GUT}=10^{16}GeV$
and $M_{SUSY}=1TeV$  and the constraints on the string models
from  the large mass hierarchy in these models.

In the following, we will first briefly review the string-inspired
supersymmetry breaking schemes before we introduce our scheme.
The mechanism of supersymmetry breaking is one of the major challenge
in extracting low-energy predictions from superstrings.  One expects that
supersymmetry breaking mechanism will bridge the huge gap between the
string theory at around the Planck scale and the standard model at around
the weak scale. Furthermore, one hopes that the non-perturbative effect, which
breaks  supersymmetry, will remove the flat-directions corresponding to
 dilaton and moduli fields, dynamically determine their vacuum
expectation values, and thus enhance the predictive power of string models.
So far the most promising
supersymmetry breaking schemes are non-perturbative.
The first attempts \cite{wit2,subr1} to break SUSY
non-perturbatively in string theory invoked gaugino condensation in the
hidden sector $E_8$. Because  dilaton couples to gauge fields,
gaugino condensation would generate a superpotential for the  dilaton. From
renormalization group invariance and dimensional analysis, the superpotential
takes the form:
\begin{equation}
W_c\,\sim \,  exp(-3S/2b_0).
\end{equation}
However in the case that the
other part of superpotential  $W_0$  is zero, the above superpotential
would yield a monotonic  potential energy,
\begin{equation}
V(z,\bar{z})\,=\,e^{K(z,\bar{z})}\,[K^{z,\bar{z}}(K_zW+W_z)(K_{\bar{z}}\bar{W}-\bar{W}_{\bar{z}})
-3|W|^2],
\end{equation}
and leads to a runaway dilaton, i.e. dilaton vacuum expectation value goes
to infinity.

The more formal derivation of the effective lagrangian with gaugino
condensation for supersymmetric Yang-Mills theory was first constructed in
\cite{vani}, and later extended to supergravity coupled to Yang-Mills theory
\cite{local,mivr4}. The target space modular invariance \cite{modu1}, which
reflects
the underlying string symmetry under duality transformations, has been
incorporated into the effective lagrangian description of gaugino condensation
\cite{mivr1,mivr2,mivr3}. The modular invariant formulation of gaugino
condensation is extended to include the formation of hidden matter field
condensation in \cite{tom1}. The more recent progress is on formulating the
phenomenon of gaugino
condensation in global and local \cite{glo,lo} super-Yang-Mills theory
with a field-dependent gauge coupling described by a linear multiplet.
But all these formulations does not shed much light on dilaton running-away
problem.

To solve the dilaton running-away problem,
it was proposed in \cite{wit2} to stabilize the dilaton by an
additional constant number in the superpotential, resulting from a VEV for
the antisymmetric tensor field strength on the internal space, i.e.
$W(S)\,\rightarrow \, W(S)+c$ with $c\epsilon_{ijk}=\langle H_{ijk}\rangle $.
The dynamics of the  induction of $\langle H_{ijk} \rangle$ by gaugino
condensation was pointed out in \cite{wit2,subr1}. Here we give  the argument
in \cite{wit2}.
It is well known that there exists in ten-dimensional supergravity a
supercovariant generalization of the field strength of the second-rank tensor
field that is essential in anomaly cancellations:
\begin{equation}
H = dB - \omega_{3Y}+\omega_{3L},
\end{equation}
with B being the two-index antisymmetric tensor field and $\omega_{3Y}$ and
$\omega_{3L}$ being  Yang-Mills and Lorentz Chern-Simons symbols:
\begin{eqnarray}
\omega_{3Y}&=&\frac{1}{30}Tr(AF -\frac{1}{3}A A A), \\
\omega_{3L}&=&\frac{1}{30}(\omega\  R -\frac{1}{3}\omega \omega \omega),
\end{eqnarray}
with Tr meaning trace in the adjoint representation and $\omega$ is the Lorentz
connection.
It was observed in \cite{wit2} that the terms in the lagrangian that depend on
$H_{ijk}$ and the gauge invariant gaugino bilinear
$\bar{\lambda} \Gamma_{ijk} \lambda$ combine into a perfect square,
\begin{equation}
\Delta L \sim (H_{ijk}-
g_{10}^2\sqrt{2}\phi^{3/4}\bar{\lambda} \Gamma_{ijk} \lambda)^2.
\end{equation}
The appearance of the perfect square means that when gaugino bilinear obtains
vacuum expectation value after gaugino condensation, dynamics forces
$H_{ijk}$ to obtain a vacuum expectation value to yield minimum energy.
One can also see how this dynamics takes place by applying the equations
of motion, as has been first pointed out in \cite{subr1}.

In \cite{wit2}, it is proposed that
$\langle H_{ijk}\rangle=c\epsilon_{ijk}=\langle dB \rangle$.
In this case, unfortunately,
c is integer  quantized in planck units \cite{rom}, so the gaugino
condensation is forced to
take place at around the Planck scale.
It is  proposed in \cite{ross}
that the necessary constant term comes from matter field VEVs. In this case,
to generate hierarchy, very small hidden sector gauge group is required.
The introduction of the modular
invariant gaugino condensation dynamics does not help solve dilaton run-away
problem \cite{mivr1,mivr2,mivr3}. To confront this problem,
scenarios in which hidden sectors have  several gauge sub  groups
or contain matter fields have been proposed \cite{kay}. The systematic analysis
of this scheme has been carried through in \cite{dixon,tom3,cas}, which show
that for pure gaugino condensation, no realistic results can emerge, even if
the hidden gauge group is not simple. And all these supersymmetry breaking
schemes do not shed much light on
the gauge coupling constant unification problem in string phenomenology.

In this letter, we propose a modified version of the mechanism proposed in
\cite{wit2,ross}. In particular, we assume that the constant term
$c\epsilon_{ijk}=\langle H_{ijk} \rangle$,
which  is turned on when gaugino condensation occurs,  comes from the
Chern-Simon term rather
than from the  anti-symmetric tensor strength or matter fields.
Since the gauge Chern-Simon part
does not have to be quantized in the Planck units as  the antisymmetric
tensor strength does \cite{rom},  we do not have the problem that
gaugino condensation has to take place  at around the Planck scale here.
It is easy to see that, in this case,  the VEV of $H_{ijk}$ can come from some
 charged background field VEVs in the compactified space. We assume
that these induced charged
background fields are in the  $E_6$ matter sector instead of in the
hidden sector ( which
except for some  phenomenological reason that we will explain later,
we can not explain why
this should be the case ). Furthermore, We require that
the induced Chern-Simon term satisfies the condition: $d\omega=F_{ij}F^{ij}=0$,
i.e.  $F_{ij}=0$. This requirement guarantees that the induced
VEVs of the charged background fields will not generate more potential
energy besides the part coming from its  contribution to superpotential.
Notice  that these charged background fields do not have to correspond to
any four-dimensional physical modes, so they can be in the adjoint
representation of the
observable gauge interaction $E_6$. As we will show later that
this fact  makes the affine level one SU(5) or
SO(10) string model with intermediate gauge symmetry breaking
to the standard model possible
in this supersymmetry breaking scheme.

What is the consequence of turning on VEVs of some
charged background fields? One direct consequence is that they will  break the
gauge symmetry through the string Higgs effect \cite{hig1}. Specifically,
the VEVs of the charged background give masses to the matter fields $\Phi$ via
the cubic  superpotential of the form $W \sim A\Phi\Phi$ and break  gauge
symmetry to the gauge subgroup that commute with the VEVs of the charged
background $A^{\alpha}$. The masses of the matter fields will be of
the order of
$\langle A^{\alpha}\rangle$. As $A^{\alpha}$ do not correspond
to any four-dimensional
physical modes or  matter fields, they can be in any representations
including the adjoint
representation. Notice that all the proofs that there is no adjoint
representation higgs fields in affine level one string models
is based on the assumption that they are four-dimensional
matter fields. Here $\langle A^{\alpha}\rangle$ are actually a special kind of
Wilson lines, and they can be in the adjoint representation. Thus in this
supersymmetry breaking scheme, level one SU(5) or SO(10) string models with
intermediate gauge symmetry breaking to the standard model is possible.
The gauge symmetry breaking scale should be on the order of
$\langle A^{\alpha} \rangle$ which is roughly the gaugino condensation
scale:
\[ \langle A \rangle \sim M_s exp(-\frac{S}{2b_0}).\]
With $\langle S\rangle =2$, for affine level one string models and
$E_6$ hidden sector gauge symmetry
 $b_0=36/16\pi^2$, we find $\langle A \rangle \sim 10^{16}GeV$
( for $E_8$ hidden sector,
$b_0=36/16\pi^2$, one gets  $\langle A \rangle \sim 10^{17}GeV$) .
This intermediate  gauge symmetry breaking scale can
correspond to $M_{GUT}$ derived in
minimal SUSY breaking grand unification models using the prescription
proposed in \cite{rulin2}.
In addition, we think that the heavy massive fields  will not
change the weak-scale predictions much, because these masses
are all on the order of $M_{GUT}$.

One can also solve the doublet-triplet problem in this supersymmetry
breaking scheme. One simply needs to require that the induced charged
background field VEVs and matter fields
satisfy a larger global symmetry and  the pseudo-Goldstone schemes in
\cite{dt} will guarantee that there are not any light triplet fields after
gaugino condensation. Here, the induced charge background
field VEVs play the exact same role as the usual Higgs fields, except that,
in the four-dimensional effective theory, they do not correspond to any
physical modes.

One would like to ask how likely this mechanism is. We think that
the induction of charged background field VEVs is more likely to happen than
that of the anti-symmetric tensor because the latter has to be quantized in
Planck units. But we do not have any dynamical argument that a gaugino bilinear
VEV in the hidden sector induces  charged background field VEVs in the
observable sector.  To break SU(5) or SO(10) string models to the Standard
model and to solve the doublet-triplet problem, some constraints are put
on the induced charged background field VEVs.
As the  existence of non-trivial charged background field VEVs
requires nontrivial topology in the compactified spacetime
( otherwise one can always gauge away such VEVs ),
to make this scheme viable, a string model is required to have some
non-trivial topology in the compactified space.
Furthermore, to obtain $M_{GUT}=10^{16}GeV$,
the hidden gauge group is required to be $E_6$ below the string scale
if one assumes $\langle S \rangle \sim 2$.
We  conclude that
in this supersymmetry breaking scheme, the derivation of $M_{GUT}=10^{16}GeV$
and the solution of the doublet-triplet problem put non-trivial constraints
on a string model just as the low-energy gauge interactions and the number of
low-energy particle generations do.

The next question to ask is how to generate the large mass hierarchy between
$M_{GUT}=10^{16}GeV$ and $M_{SUSY}=1 TeV$ in this scheme. Obviously,
after  gaugino condensation occurs, the gravitino gets a mass $M_{3/2} \sim
\Lambda ^3/M_P^2 \sim 10^{12} GeV$.
If global supersymmetry in the matter sector is broken at about the same
scale as one normally assumes, one can not generate the large mass hierarchy
between $M_{GUT}=10^{16}GeV$ and $M_{SUSY}=1TeV$ in this scheme.
To confront this problem, we consider a special kind of string
models in which moduli and matter fields are of a no-scale structure
\cite{noscale}, which is
shown to be a generic structure for a class of string models\cite{wdr,rara}.
It has been shown that for no-scale  models, when local supersymmetry
is broken,  global supersymmetry is still preserved \cite{no}.
To be more specific, in the most simplistic case, the K\"ahler
potential takes the
form:
\begin{eqnarray}
K & = &-\ln(S+\bar{S})-3\ln(T+\bar{T}-|\phi|^2), \\
W & = & h e^{-3S/2b_0} + W_l +W_0.
\end{eqnarray}
Here $\phi$ are  matter fields, S is dilaton field, and T is moduli field, and
h is some constant on the order of one. $W_l$ is
the superpotential coming from the
induced charged background VEVs \footnote{One would ask whether the induced
$W_l$ terms would break modular invariance. In fact, modular
invariance is only broken by $\langle A^{\alpha}\rangle$ spontaneously.
As ten-dimensional fields, $A^{\alpha}$ transform under modular transformation:
$A^{\alpha}\rightarrow A'^{\alpha}=\frac{A^{\alpha}}{icT+d}$
and $W_l \rightarrow W'_l=\frac{W_l}{(icT+d)^3}$,
modular invariance is preserved.}
 and $W_0$ is the superpotential for matter
fields.
In string models, moduli correspond to some flat-directions; $W_0$ does not
depend on moduli explicitly.
The non-perturbative effect, gaugino condensation, may generate
a moduli-dependent superpotential
\cite{mivr1,mivr2,mivr3}
for the string models \cite{kap,dixon2} in which the gauge coupling
function receives moduli-dependent
string threshold corrections.
It has been shown that \cite{toms} for string models with no sectors
preserving N=2 spacetime supersymmetry, for example, $Z_3,Z_7$
orbifold models,
 the gauge coupling constant does not receive  moduli-dependent string
threshold correction. In this type of string
models, one obtains a moduli-independent  non-perturbative superpotential
which is one
of the crucial ingredient for being no-scale models. As we will show later,
a moduli-independent
gauge coupling function is  also a necessary requirement for generating small
 observable sector gaugino
masses (as compared to $10^{16}GeV$). So in our supersymmetry breaking scheme,
 the large mass hierarchy restricts  string
models to  this type only.
 Assuming that the superpotential
does not depend on moduli, we find that
the vacuum potential energy takes the following form:
\begin{eqnarray}
V & = & e^K ( |(S+\bar{S})W_S-W|^2 + \sum_i|W_{0i}|^2 \nonumber ) \\
  & = & e^K|[1+\frac{3}{2b_0}(S+\bar{S})]
exp(-3S/2b_0)+W_l+W_0|^2 + \sum_i |W_{0i}|^2.
\end{eqnarray}
We see that  the potential energy is  positive semi-definite
 ( so that  the cosmological constant is naturally zero ) and vanishes at
\begin{equation}
(S+\bar{S})W_S-W=0,
\end{equation}
or
\begin{equation}
 [1+\frac{3}{2b_0}(S+\bar{S})]\,hM_s^3 exp(-3S/2b_0)+W_l+W_0=0,
\end{equation}
and
\begin{equation}
W_{0i}=\frac{\partial W_0}{\partial \phi_i}=0.
\end{equation}
For $\langle W_0\rangle =0$, we  have
\begin{equation}
 [1+\frac{3}{2b_0}(S+\bar{S})]\,hM_s^3 exp(-3S/2b_0)+ W_l=0.
\end{equation}
It is  easy to see that the solution to the above equations yields
nonzero $\langle W \rangle$.
This will give the gravitino a mass and break  local supersymmetry
while the cosmological constant remains zero.
A simple calculation shows that  supersymmetry is broken in the dilaton
and moduli sector; but in the observable matter sector,  scalar masses and
A-terms remain zero. We restrict the string models to those in which the gauge
coupling function
does not depend on  moduli fields. In this case, at tree level, the
gaugino mass is zero ( otherwise the gaugino mass will be too large )
and global supersymmetry is preserved in the matter sector.
We conclude that, in this supersymmetry breaking scheme,
the large mass hierarchy constrains string  models to be those
that have a no-scale structure and in which the gauge coupling constant does
not
receive string threshold corrections.

We hope that for some string models, radiative corrections may break
the preserved global supersymmetry at low-energy and the large mass hierarchy
between $M_{GUT}=10^{16}GeV$ and  $M_{SUSY}=1 TeV$ could be generated.
To accomplish this, we are facing a nontrivial problem which is
closely related to that of finding
the true vacuum to one-loop order and determining dilaton and moduli VEVs.
It is easy to see that in this kind of supersymmetry breaking models,
all the mass scales involved are determined by the dilaton and moduli VEVs,
which
unfortunately correspond to some flat directions at tree level and can not be
specified. These flat directions could be removed  by loop-effects and dilaton
and moduli VEVs can be dynamically
determined by radiative-corrected potential energy.
But with the K\"ahler potential
and superpotential  proposed here, we can not get the desired dilaton VEV
from the one-loop effective potential.
Another problem with the above formulation is that it is not modular invariant
in the dilaton sector.
Considering the complexity of the problem, we defer a more detailed discussion
to future work.

In conclusion, we have  proposed that, instead of an antisymmetric
tensor field \cite{wit2,ross} or matter fields \cite{ross} getting VEVs,
some charged background field VEVs might be induced when gauginos condense
in the hidden sector. The advantages
of this kind of supersymmetry breaking models is that one does not
have to appeal to string
threshold corrections, higher affine level string models, or extra
heavy fermions  to solve the gauge
coupling constant unification problem in string phenomenology.
 It connects the gaugino condensation scale with
the local supersymmetry breaking scale and $M_{GUT}$ and generates
these scales dynamically. It makes the affine level one SU(5) or SO(10) string
models with  intermediate gauge symmetry breaking to the standard model at
 $M_{GUT}=10^{16}GeV$ possible; the doublet-triplet problem can also be solved
in this scheme. We show that the solution of these problems puts non-trivial
constraints on a string model. For example, the
experimentally implied result, $M_{GUT}=10^{16}GeV$, constrains the hidden
sector gauge interaction below the string scale to be $E_6$ if one assumes
 $\langle S \rangle \sim 2$.
 To generate the large mass hierarchy in this scheme, we show
 that string models are constrained to be those that have a no-scale
structure and in which the the gauge coupling constant does not receive string
threshold correction. We conclude that the derivation of $M_{GUT}=10^{16}GeV$,
the  solution of the doublet-triplet and the large mass hierarchy problem,
 put non-trivial constraints on a string model.
Compared to the constraints
on a string model from low-energy gauge interactions and fundamental particle
generations, the constraints from the large mass hierarchy and
$M_{GUT}=10^{16}GeV$ can be thought as
dynamical constraints, which in the schemes proposed here, put much more
direct and restrictive constraints on string models.

\vskip 28pt
\noindent{\bf Acknowledgement}
\vskip 12pt

I would like to thank Mary K. Gaillard, Steve Kelley,
Dick Arnowitt and Jim Liu for inspiring discussions.
\vskip 28pt

\vfill\eject

\end{document}